\documentclass[runningheads]{llncs}
\usepackage{graphicx}
\usepackage[llncs]{llncsconf}
\conference{}
\llncs{This is a post-peer-review, pre-copyedit version of an article published in Lecture Notes in Computer Science (LNCS, volume 11842).}{1}
\llncsdoi{10.1007/978-3-030-32872-6_15}

\begin{document}
\title{SQUADfps: Integrated model-based machine safety and product quality for flexible production systems}
%
\titlerunning{SQUADfps for machine safety and product quality}
%

\author{Chee Hung Koo\inst{1} \and
Stefan Rothbauer\inst{2} \and
Marian Vorderer\inst{1} \and
Kai H{\"o}fig\inst{2,3} \and
Marc Zeller\inst{2}}
\authorrunning{C.H. Koo, S. Rothbauer et al.}
%
\institute{Robert Bosch GmbH, Corporate Sector Research and Advance Engineering,\\ 71272 Renningen, Germany\\
\email{\{givenname.surname\}@de.bosch.com}
\and
Siemens AG, Corporate Technology, Otto-Hahn-Ring 6, 81739 Munich, Germany\\
\email{\{givenname.surname\}@siemens.com}
\and
University of Applied Science Rosenheim, Hochschulstrasse 1,\\ 83024 Rosenheim, Germany\\
\email{kai.hoefig@fh-rosenheim.de}
}
\maketitle              
\begin{abstract}
Growing individualization of products up to lot-size-1 and high volatility of product mixes lead to new challenges in the manufacturing domain, including the need for frequent reconfiguration of the system and reacting to changing orders. Thus, apart from functional aspects, safety aspects of the production system as well as product quality assurance aspects must be addressed for flexible and reconfigurable manufacturing systems at runtime. To cope with the mentioned challenges, we present an integrated model-based approach SQUADfps (machine \emph{S}afety and product \emph{QUA}lity for \emph{f}lexible \emph{p}ro\emph{D}uction \emph{s}ystems) to support the automatic conduct of the risk assessment of flexible production scenarios in terms of safety as well as the process-FMEA to ensure that the requirements w.r.t. the quality of the production process and the resulting product are met. Our approach is based on a meta-model which captures all information needed to conduct both risk assessment and process-FMEA dynamically during the runtime, and thus enables flexible manufacturing scenarios with frequent changes of the production system and orders up to a lot-size of one while guaranteeing safety and product quality requirements. The automatically generated results will assist human in making further decisions. To demonstrate the feasibility of our approach, we apply it to a case study.
\end{abstract}
\section{Introduction}
%
Major trends in the manufacturing sector are the growing individualization of products and volatility of product mixes. If taken to extremes, this scenario also counts for products being produced only one time (lot-size-1) or only on demand.
In order to reach this goal, the concept of \emph{Flexible Manufacturing Systems (FMS)}, which can change their software during runtime \cite{YILMAZ1987209}, and \emph{Reconfigurable Manufacturing Systems (RMS)}, which can adapt their software as well as their hardware \cite{KOREN1999527}, play a vital role. Moreover, standalone systems from different manufacturers are interconnected to accomplish a common production goal and the production processes can be orchestrated automatically in so-called \emph{Plug-and-Produce} scenarios \cite{ARAI20001}. 

Due to frequent changes of the products being manufactured, the rapid adjustment of a factory is a major challenge to implement application scenarios of flexible production systems (often called Industry 4.0 or Cyber-Physical Production Systems) successfully. Although the high flexibility of future flexible production scenarios promises a faster market adaptation and responsiveness, it raises at the same time dependability-related concerns due to unknown configurations at runtime. Thus, apart from functional aspect (i.e.~the check if a factory is able to manufacture a specific product), safety aspects as well as product quality assurance aspects must be addressed.

Safety standards, such as ISO 13849 \cite{iso13849} or IEC 62061 \cite{iec62061} in context of industrial production systems, provide guidelines to keep the residual risks in machine operation within tolerable limits. For every production system, a comprehensive risk assessment is performed, which includes risk reduction measures if required (e.g.~by introducing specific risk protective measures such as fences). The resulting safety documentation describes the assessment principles and the resulting measures that are implemented to minimize hazards. This documentation lays the foundation for the safe operation of a machine and it proves the compliance with the Machinery Directive 2006/42/EC of the European Commission \cite{directive}.
In flexible production scenarios, risk assessment must be conducted after each reconfiguration of the production system. Since this is a prerequisite for operating the factory in the new configuration, a manual approach can no longer effectively fulfill the objectives for assuring the safety in highly flexible manufacturing scenarios. Hence, the acquisition of safety-related information from each individual production step and the analysis of possible emergent hazards must be conducted in an automated way to quickly assess a new configuration of a manufacturing plant.

To evaluate the quality of a product considering the production process, a \emph{Process Failure Mode and Effects Analysis} (process-FMEA) is typically performed. During production, every process step can negatively influence the quality of the product depending on the negative outcome of the process step. This is captured in a process-FMEA with the concept of failure modes of a process step as well as the respective severity. It also defines measures to detect and deal with unwanted effects on product quality. Such an analysis is important to document the applied quality measures and to find out where drawbacks in the production process are and how they can be addressed.
Since the factory's configuration as well as its products constantly change in adaptable and flexible factory scenarios, a process-FMEA must be performed dynamically during the production of each product based on the configuration used. This is necessary to ensure that the products requirements w.r.t.~quality will be met by the provided production process.

In this paper, we present an approach for the model-based assessment of flexible and reconfigurable manufacturing systems based on a meta-model. This integrated approach SQUADfps (machine \emph{S}afety and product \emph{QUA}lity for \emph{f}lexible \emph{p}ro\emph{D}uction \emph{s}ystems) captures all information needed to conduct both risk assessment and process-FMEA dynamically during the runtime of the manufacturing system in an automated way. In this way, the approach enables flexible manufacturing scenarios with frequent changes of the production system up to a lot-size of one. In order to provide a better understanding for our proposed approach, we assume that the considered production systems are already installed as intended and focus only on the reconfigurability in terms of equipment and process changes.  

The rest of the paper is organized as follows: 
In Chapter \ref{relatedwork}, an overview of related work is given. Chapter \ref{modelbasedsafetyassessment} introduces a meta-model (Section \ref{metamodel}) that enables an automated hazard and risk assessment for the factory (Section \ref{modelbasedriskassessment}) and a process-FMEA for a product to be produced in a factory to ensure that the its quality is on track (Section \ref{dynamicprocessFMEA}).
Chapter \ref{casestudy} presents a case study to show how the model can be applied.
Chapter \ref{conclusion} summarizes the main results and provides an outlook for further research topics.

\section{Related Work}
\label{relatedwork}
The usage of model-based approach to carry out safety analyses or safety assessment aims to achieve compositional, reusable safety assessment and to improve traceability of information provided from the system design phase \cite{lisagor2011model}. The collective term \textit{model-based safety assessment} includes a wide range of techniques proposed in the academia \cite{joshi2005model,lisagor2011model} that have already been applied extensively nowadays in varying domains such as the automotive industry \cite{PAPADOPOULOS200577}, IT security \cite{houmb2002towards,1137696}, aviation sector \cite{bernard2007experiments}, train protection system \cite{wang2009study} and industrial automation e.g. collaborative robots application \cite{8247648}. In most of these applications, the safety requirements of the designed systems are assessed based on the functional system models created. Different tools and modeling techniques have also been developed since then to facilitate and maintain model-based safety engineering and safety analysis \cite{cancila2009sophia,grigoleit2016qsafe,prosvirnova2013altarica}. However, most of the mentioned publications deal with model-based applications during design and development phases instead of \textit{runtime} applications, which is one of the most important aspects for highly flexible manufacturing scenarios. To facilitate flexible and reconfigurable manufacturing systems in a practical way, as safety analyses nowadays are an inherently manual tasks in which only very few steps are automated, it is necessary to support these manual processes with automation as far as possible. Frequent system changes needed for lot-size-1 scenarios require runtime safety assessment to be done in an economically feasible manner\cite{koo2018challenges}. In this paper, we propose a method to carry out safety assessment automatically at runtime using a proposed meta-model, which can facilitate human during the decision-making to approve new system configuration. 

\textit{Failure mode and effects analysis} (FMEA) has its origins in military applications \cite{MIL-STD-1629A} and was used in the same decade to analyze the influence of failures in production processes \cite{FordFMEA}. Since it is an effective but costly analysis technique, automating it has a long history in functional safety \cite{cichocki00G,cichocki01G,david.2008,papadopoulos04,walker.2009.a} and also for analyzing machinery. 
In \cite{IntegratedFMEA}, it is mentioned that process-FMEA is part of an integrated approach for safe products, but that classic process-FMEA does not consider the manufacturability of a product influenced by quality problems. In \cite{CPP}, the authors use FMEA among other techniques to assess the manufacturability and estimating the cost of a conceptual design in early product design phases. They introduce an approach to estimate costs of failures during manufacturing using an extended FMEA approach introduced in \cite{FMERA}. This is a manual task that is used to prioritize different manufacturing options. Their work can be used in combination with the approach presented here to include costs of potential failures during manufacturing. In \cite{PPR}, product process resource-based approach is presented that uses an ontology to model the manufacturing capabilities and the required process steps to produce a product. Similar to the approach presented here, the authors use a standardized language set in an ontology to (semi-)automate the process of mapping production steps for a product to machinery. Nevertheless, they do neither aim for quality aspects of the output nor for rejected items in the mapping process. 

\section{Model-based Safety and Quality Assessment of Flexible, Adaptable Production Systems}
\label{modelbasedsafetyassessment}
\subsection{Meta-model}
\label{metamodel}
\begin{figure*}[ht!]
\centerline{
\includegraphics[scale=0.5]{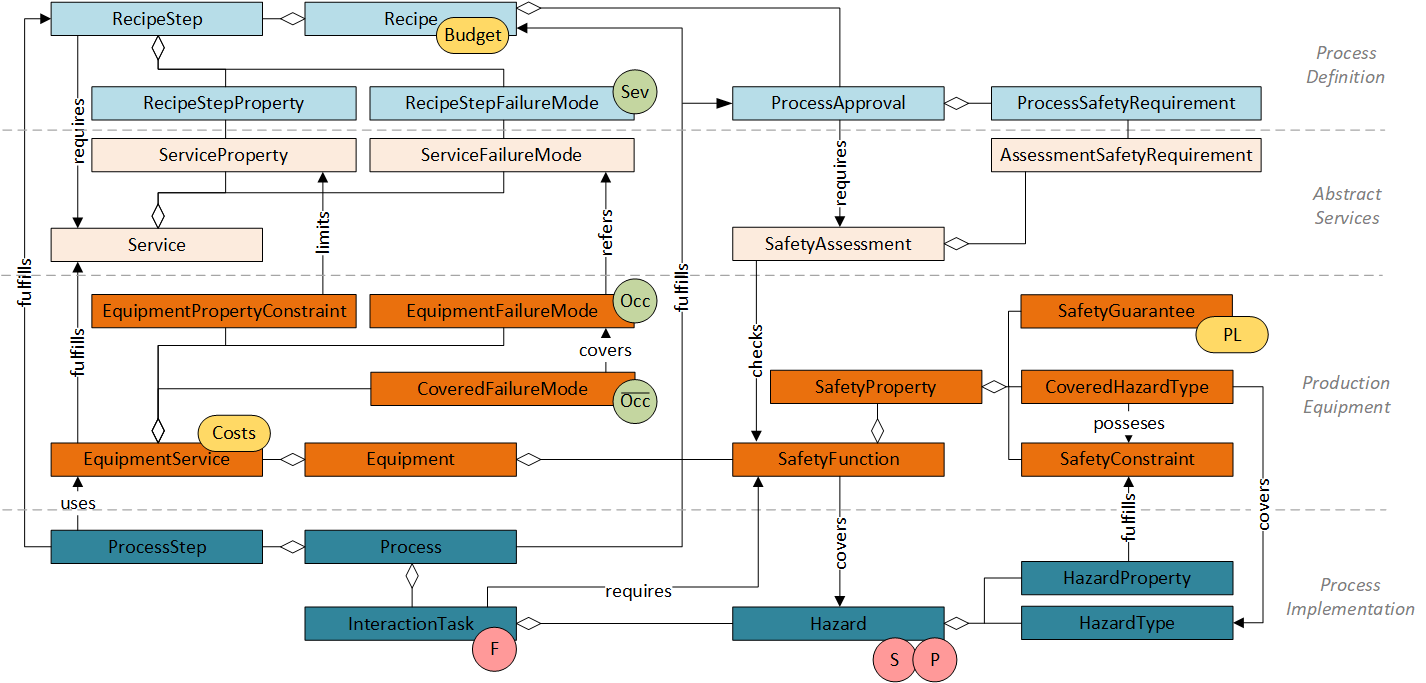}
}
\caption{SQUADfps meta-model for process-FMEA and safety assessment supporting automation}
\label{fig.pfmea}
\end{figure*}

Figure \ref{fig.pfmea} shows the proposed meta-model of SQUADfps for a flexible and reconfigurable production system. The meta-model is divided into four categories, considering both machine safety and product quality aspects:

\begin{itemize}
\item{\textbf{Process Definition}: In the product category, the elements address the order and steps related to \emph{what} has to be done to produce a product. This category also addresses the required safety approval process before the production is allowed to commence.}
\item{\textbf{Abstract Services}: The model elements of the category abstract services collect common specification of services and service parameters across all factories. These elements enable the specification of a product independently from a concrete production equipment. Besides, this category provides abstract service to carry out safety assessment for any concrete production equipment.}
\item{\textbf{Production Equipment}: The elements of this category model a concrete factory or production system along with its machinery equipment, describing what it can do, what quality measures are available and what safety functions are implemented.}
\item{\textbf{Process Implementation}: In the process implementation category, the elements address the concrete process used to produce a product. Here, the process steps address concrete ordered actions that are executed to produce a product. These steps provide a solution on \emph{how} a product is produced. Besides, concrete hazards associated to the process are identified.}
\end{itemize}

These categories allow users to map different activities, use cases and roles in the domain of dynamic reconfigurable production scenarios to automatically generate a process-FMEA (quality of the product) and a risk assessment result (safety of the production system) for the production system under consideration. 

In the \emph{process definition} category, the product owner specifies \emph{which} steps are needed and in which order they need to be executed to produce a product (recipe). The product owner addresses abstract services that can satisfy steps of its recipe. Those abstract services provide a global library of all available services. Each service declaration can have constraints and parameters that can be set for a recipe step (service property). For example, the abstract service \emph{drill} requires the rotation speed of the drill and the size of the drill hole as parameters. When the abstract service is instantiated in a recipe step, these parameters need to be set.

Different failure modes can be stored (failure mode declarations) during the abstract service declaration. Independent from the concrete equipment or machinery (equipment), failure modes are known and defined in general. For example, the service \emph{drill} has the potential failure mode \emph{skew drill hole} for all concrete machinery implementing this service. 

For each addressed service declaration in a recipe step, the failure mode declarations are known to the product owner that defined the recipe. He now can specify how severe the different failure modes (using recipe step failure modes) are for his product. Thus, the first step for the quality assessment using a process-FMEA can be performed without knowledge of the concrete equipment that later produces the product. This can be done for the combination of recipe steps and failure modes rated with a severity value. 

Independent from this specification scenario of a recipe, the owner of a factory can model the equipment (\emph{production equipment}) with equipment services and safety functions. Equipment services address the abstract service declarations available in the global library. Equipment property constraints are used to specify the possible operating parameters and limitations of service property declarations, while equipment failure modes address the service failure modes of the abstract service. 

During declaration of \emph{production equipment}, the factory owner can specify the available machinery and the equipment service along with its parameter limitations, which can be provided for a specific recipe. With this, the factory owner gets a list of possible abstract failure modes and can specify how often the abstract failure mode occurs for the concrete service (equipment failure mode). This can be known either by previously collected data or data provided by the manufacturer of the machinery. In this case, the factory owner can provide information about the occurrence value of concrete failure modes while using the equipment during the production.

In order to consider the safety of the production, the process will require safety approval before operation (process approval) during \emph{process definition}. Process safety requirements specify the minimal safety requirements to be achieved. For instance, the product owner can specify that only safety functions with a certain minimal safety guarantee are allowed due to safety criticality of the product or enforced safety guidelines. This process approval addresses the relevant abstract service (safety assessment), which checks whether all expected hazards are covered by the available safety functions considering the risk level. 

Beside modeling the failure model for process-FMEA, the factory owner can also model the safety functions provided by an equipment. For instance, an equipment protective measure such as light curtain that is installed can protect the personnel during interaction with the equipment, which provides safety guarantee in term of \emph{performance level} to describe the reliability of the safety device. A safety function covers certain hazard types, which can be described through predefined semantics. A light curtain can protect personnel against mechanical hazards (crushing, shearing etc.), as long as the hazard source lies within the allowed working area and occurs during certain interaction tasks (safety constraint).

During \emph{process implementation}, the factory owner will get a list of possible hazards in relation to the interaction tasks, in which the frequency of the task can be defined. The risk parameter frequency describes the interaction of personnel with the production system. For a lot-size-1 scenario, the frequency can still be defined as high if the responsible personnel needs to carry out manual tasks for a foreseeable high amount of time. In combination with the concrete risk parameters (severity and possibility for avoidance) provided by the equipment, an identified hazard with its evaluated risk level (hazard property) can be checked against the safety function to ensure the production safety. Further examples for hazard properties include runtime location of hazard source, moving speed of its equipment, relevant interaction tasks etc.

\subsection{Model-based Risk Assessment}
\label{modelbasedriskassessment}
As mentioned before, a production process might include some human interaction tasks in different life cycle phases, such as setup of equipment, interactions during the production or maintenance activities. These interactions are specific to the concrete process and independent from the recipe, which describes the product to be manufactured. Each interaction task can include one or more hazards for the personnel involved, which have a certain level of severity. Each hazard also possesses a possibility of avoidance, which determine how possible a person can avoid the hazard during its occurrence. According to the risk graph in the standard ISO 13849 \cite{iso13849}, the risk level of a particular hazard can be evaluated using the severity of associated hazard (S), the frequency of tasks (F) and the possibility of avoidance (P). This leads to a risk level described in term of \emph{required performance level} (PLr).

Safety functions are typically installed to protect humans against a certain hazard and have a \emph{performance level} (PL) value, which describes the overall reliability of the safety device considering the components used. Having this information provided by the machine vendors, the required performance level gained from risk assessment (PLr) can be evaluated against the provided safety function performance level (PL). In a conservative manner, the production process can only be approved manually by the factory owner when all the identified risks are covered successfully by the available safety functions considering PL value. 

\subsection{Dynamic process-FMEA}
\label{dynamicprocessFMEA}
Since equipment is not only able to execute production steps in a recipe, but is also able to execute quality measures, an equipment service can therefore cover certain failure modes. These measures can be originating from the same service, from a different service of the same equipment or from a service of a different equipment. For example, a robot arm that can be used to perform pick and place can also supervise its own actions using a camera. In this case, the failure mode \emph{misplacement} of the service \emph{pick and place} can be covered by the service \emph{camera supervision} from the same equipment. Using this methodology, the factory owner can specify which machinery can be used to increase the quality of the production. Since quality measures decrease the occurrence of certain failure modes, each covered failure mode stores a decreased occurrence value. 

Using the severity of a failure mode from the product specification (recipe failure mode) multiplied by the occurrence value of the equipment failure mode or with the decreased occurrence value of a quality measure, a process-FMEA can be conducted for a product produced by a certain process on a concrete set of equipment. 

This model-based approach ensures a structured and systematic analysis for all known failure modes that are captured within the model. This is valuable, as systematic and complete analysis is a requirement e. g. required by safety or quality standards. If experience from production about failures that actually were observed but not yet captured in the model is included, over time the analyis should become complete with regard to present knowledge. During the first applications in real production there might be a need to at least verify completeness by a manual inspection. A manual inspection and possibly extension of a pre-generated pFMEA is much less effort than starting from scratch, so even at the introductory phase there is already a reduction in effort to be expected.

\section{Case Study}
\label{casestudy}

\begin{figure}[ht]
\centerline{
\includegraphics[scale=0.86]{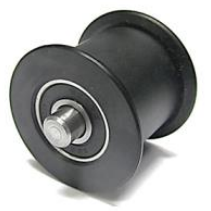}
}
\caption{Roll with axle}
\label{fig.rolle}
\end{figure}

\begin{figure}[ht]
\centerline{
\includegraphics[scale=0.86]{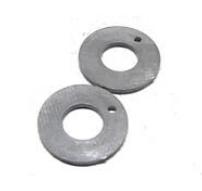}
}
\caption{Disc to be mounted and greased}
\label{fig.scheibe}
\end{figure}

\begin{figure}[ht]
\centerline{
\includegraphics[scale=0.8]{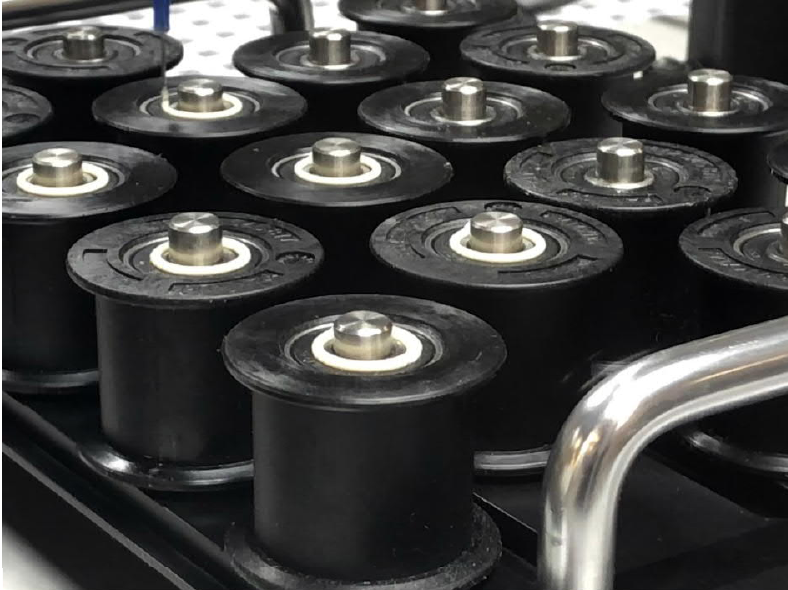}
}
\caption{Tray with products}
\label{fig.tray}
\end{figure}

In this small case study, we want to demonstrate how to use the meta-model as described in Section \ref{modelbasedsafetyassessment}. The product that we investigate here is a small roll that consist of a roll body, an axle and two metal discs as depicted in Figure \ref{fig.rolle} and Figure \ref{fig.scheibe}. The entire material is delivered on a tray, see Figure \ref{fig.tray}, and is set together by a robot arm that also greases the contact area of the parts. After that, a visual inspection detects insufficient products. To rate failure modes, we use an risk priority number (RPN) based approach for the parameters \emph{severity (sev)}, \emph{occurrence (occ)} and \emph{detection (det)} with a range from one to five whereas one represents the lowest severity, a negligible occurrence rate and a sure detection and five represents a high severity, a high occurrence rate and an nearly impossible detection. For the assessment of machine safety for the setup production system, performance level (PL) is used in accordance with ISO 13849-1\cite{iso13849}. Risk level of the identified hazards can be described through \emph{required} performance level (PLr), with the risk parameters \emph{severity}, \emph{frequency} and \emph{possibility of avoidance}, whereas \emph{PL a} represents the lowest tolerable risk and \emph{PL e} represents the highest risk.

\subsection*{Dynamic process-FMEA}
The recipe steps for production are depicted on the left side in table \ref{fig.example_neu} for recipe $R=r_1,\dots,r_6$. For the first process $P=p_1,\dots,p_{6a}$ , the tray is delivered using an abstract service \emph{convey} which is implemented by the equipment \emph{belt conveyor}. The failure modes of this service are \emph{misplacement} and \emph{shock} rated by the design team with a severity value of four and five respectively. The production equipment produces failures with a occurrence of two and one. A visual inspection can safely detect both failure modes (detection value \emph{Det} is 1). 

\begin{table*}[ht]
\caption{Example product recipe and two processes using abstract services.}
\centerline{
\includegraphics[scale=0.6]{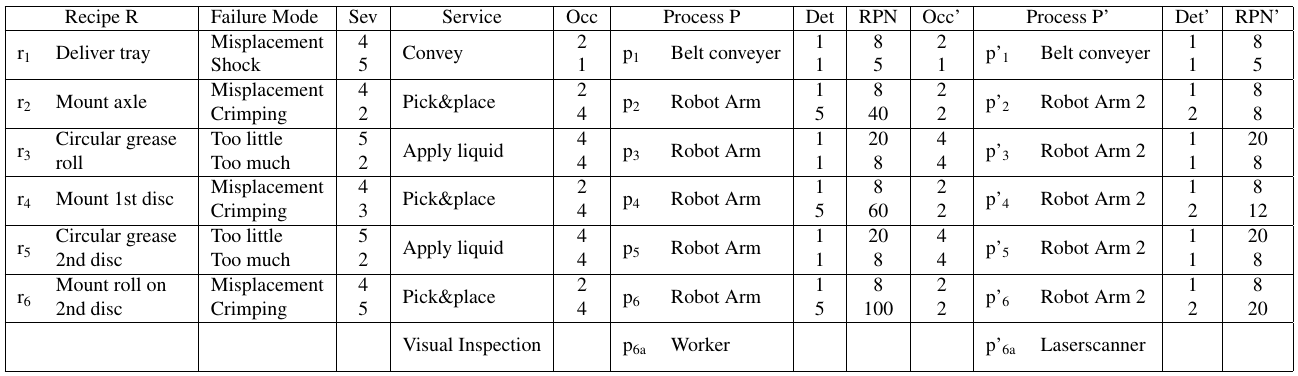}
}
\label{fig.example_neu}
\end{table*}

The next step is to mount the axle inside the roll. This step is fulfilled by the service \emph{pick and place} which is implemented by a robot arm. This service can fail in two ways, the object can be misplaced but can also be crimped by the clutch. Crimping is not very severe to the axle since it is made from solid metal. This cannot be detected by a visual inspection (detection value \emph{Det} is 5). Both discs need to be greased and there can be too much and too little grease. Having too little grease is quite severe, and the worker can detect it. Having too much grease is just a minor failure. Since the roll itself is made from plastic material, crimping is severe since the roll can be damaged. This failure mode can hardly be detected by the worker, since he is not doing a stress test (detection value \emph{Det} is 5). 
The elements of properties and constraints are not depicted in the table for the reason of space limitations. Service properties of \emph{pick and place} would include, for example, start- and endpoint, trajectory and weight, whereas an equipment implementing this service provides limitations of those parameters and recipe steps requesting the service would provide the required information to fulfill the step. 

With the failure mode information provided by the service definition, the design team can specify \emph{what} failure mode is severe (requirement) and the vendor can specify \emph{how} often the failure mode appears on its machinery and \emph{how} the effect of the failure mode can be prevented in later products (implementation). The process $P$ generally is capable to implement the recipe $R$ since the equipment fulfills the required service of each recipe step and the relative order of the process steps matches the order of the recipe steps with an additional step at the end of the process: $p_{6a}$. 

Also depicted in table \ref{fig.example_neu} is an additional process $P'=p'_1,\dots,p'_{6a}$ that also fulfills recipe $R$ but with different equipment. A different robot arm is used, that has a lower probability of crimping. Additionally, the visual inspection is implemented by a more precise laser scanner that better detects crimping. With these two adoptions in place, the highest risk priority number is lowered from 100 to 20. 

This approach in its basic implementation is of a qualitative nature. It therefore enables comparing different production alternatives and facilitates the selection of a appropriate schedule selection for the production of a concrete product.
In a specific domain the quality criteria might be specified in a quantitative manner, failure probabilities replacing occurence values and actual costs at risk replacing severity values. If this is possible for a certain domain or use case then for products and production quantitative goals can be specified and the selection or ranking of different schedules with regard to fulfilling quality requirements of a product can be done. A manual selection will probably still be necessary to balance e. g. quality goals with other goals not captured within this model.

\subsection*{Model-based Risk Assessment}
Using the same production process described above, an example for the conduct of safety assessment using an abstract service, as depicted in figure \ref{fig.pfmea}, can also be shown. In this production process, the operator is required to load the product parts (roll body with axle) in a frequent manner onto the belt conveyor. Besides, the robot's handling tool needs to be adjusted and maintained occasionally to ensure its high precision. Hence, the task frequency of these two interaction tasks can be described as F2 (high frequent) and F1 (low frequent) respectively. As the frequency is defined in relation to the overall process duration required, its definition is hence independent from the product lot sizes.

The initial production system with \emph{Belt conveyor} and \emph{Robot Arm} in table \ref{fig.case_study_1} introduces three different hazards for the described interaction tasks. During the loading of production parts, the movement of robot arm can cause shearing points with high severity (S2), which can hardly be avoided due to its high movement speed (P2). On the other hand, the moving belt conveyor introduces possible squeezing points to the operator with the risk parameters S1 and P1 thanks to its relatively well-considered inherent design. During the maintenance of robot arm, the operator might still be bruised by the arm movement (S2) although the movement speed is monitored by its safety control function (P1). Here, only a light curtain is provided as a safety function with the performance level PL d. 

As shown in table \ref{fig.case_study_1}, the current setup does not fulfill all the safety requirements. Hazard $h_1$ with a high level of risk PL e does not receive a suitable safety function that fulfill the required performance level. In addition, there is no available safety function that can counter the hazard $h_3$. Based on this result, the responsible individual can decide whether to reduce the risks, to eliminate the risks or to provide extra protective devices to the system. This involves creative decision-making process and is not being considered in the proposed meta-model. The generated risk assessment result can then provide instant updated information after every system modification to assist the decision-making process. 

It is assumed that the financial situation allows the factory owner to acquire new equipment. In order to improve the safety of the production system, a different robot arm (\emph{Robot Arm 2}) that provides the same services is now used, as depicted in table \ref{fig.case_study_2}. This robot arm is equipped with an integrated sensor skin (\emph{safety sensitive cover}) that can detect human approaches and turn off the robot once the operator violates the safety distance. This sensor skin provides a safety assurance of PL e. With this new robot arm, all previously unfulfilled safety requirements are now satisfied by the provided safety functions. The abstract service now confirms the results and awaits safety engineer to make the final approval.

\begin{table*}[ht]

\caption{Exemplary risk assessment for the provided product recipe using abstract services (safety requirements are not completely fulfilled).}
\centerline{
\includegraphics[scale=0.4]{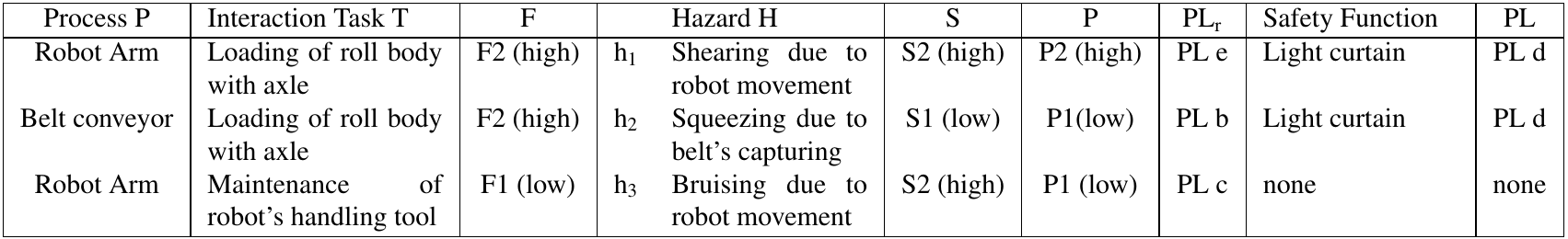}
}
\label{fig.case_study_1}
\end{table*}

\begin{table*}[ht]

\caption{Risk assessment after implementing counter measures using a different robot arm (safety requirements are now fulfilled).}
\centerline{
\includegraphics[scale=0.4]{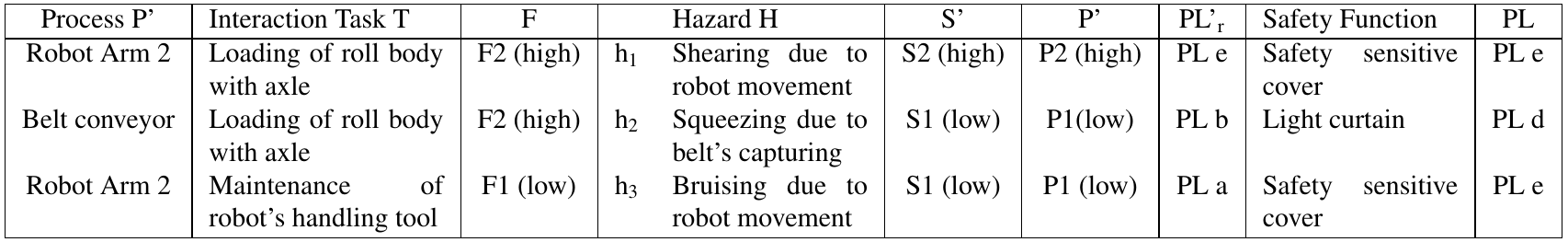}
}
\label{fig.case_study_2}
\end{table*}

This example shows how the usage of an abstract service allows the definition of an abstract production recipe without addressing concrete production equipment. The product design team uses abstract service declarations and properties to formulate production requirements. It can be decided (semi-)automatically if the production equipment can manufacture a product defined by a recipe. By providing information about the severity of certain failure modes, those requirements are extended by quality requirements. In a second step, a factory can map its production equipment to this abstract language and evaluate if it can produce the recipe. By providing information about the occurrence of failure modes of the existing production equipment, it can be evaluated using RPNs if the required quality can be met or if additional quality measures need to be implemented to increase the quality. By having a budget for a recipe, the vendor of a product can evaluate the economic efficiency of its possible production scenarios and decide to produce a product or to decline an offer. By comparing the RPNs of prospective processes and their economic deficiencies, an optimal process can be selected.

The same applies to the risk assessment procedure. By using abstract service definitions, the integrated production equipment can be checked automatically during runtime to guarantee the safety of operators while interacting with the production system. The known interaction tasks are firstly associated with information regarding possible involved hazards, whereas the severity and possibility of avoidance are then described concretely by the integrated production equipment. The frequency of interaction tasks can also be predefined in order to evaluate the required risk level using performance level PLr along with the other risk parameters. Considering the available safety functions along with its constraints at runtime, the production system can be assessed against the identified hazards, emphasizing hence the critical points that require further safety considerations and safety measures. This ensures a higher efficiency, quality and completeness of the risk assessment result, which is usually done manually nowadays.

\section{Conclusion \& Future Work}
\label{conclusion}
In this publication, we presented an integrated model-based approach SQUADfps that enables both the automated conduct of risk assessment and the dynamic creation of a process-FMEA for a flexible, adaptable or reconfigurable production system. Our proposed meta-model provides the foundation to enable flexible production scenarios in which individual and customer-specific productions can be manufactured up to lot-size-1. The proposed model-based risk assessment can ensure the safe operation of a new, previously unknown configuration of the manufacturing system by conducting the required risk assessment in an automated way based on the information available in the meta-model. Moreover, the evaluation on whether a specific product can be manufactured while meeting the customer's quality requirements by a specific configuration of the plant (as well as a cost-efficient use of quality assurance mechanisms within the manufacturing process) can be conducted by generating a process-FMEA in an automated manner. By applying the proposed model-based approach, all information required to perform these assessments can be provided automatically during runtime. The currently manual and time-consuming tasks to conduct assessments can be automated. This assists the decision-making process of human and thus, enables the fast reconfiguration of production systems in flexible production scenarios. In the future, this integrated model-based approach will be applied to further use cases to improve the completeness and significance of the generated results.

\section*{Acknowledgement}
\label{acknowledgement}
The work leading to this paper was funded by the German Federal Ministry of Education and Research under grant number 01IS16043Q and 01IS16043O (CrESt).

%
%
%
\bibliographystyle{splncs04}
\bibliography{references}
\end{document}